\documentstyle [12pt]{article}

\input epsf
\begin{document}

\begin{titlepage}

\nopagebreak

\vglue 1.5  true cm
\begin{center}
{\Large \bf
Q-States Potts model \\
\medskip
on a random planar lattice }
\vglue 1.5 true cm
{\bf Jean-Marc DAUL }

\medskip
{Laboratoire de Physique Th\'eorique de
l'Ecole Normale Sup\'erieure\footnote{Unit\'e Propre du
Centre National de la Recherche Scientifique,
associ\'ee \`a l'\'Ecole Normale Sup\'erieure et \`a
l'Universit\'e
de Paris-Sud.

\noindent
e-mail address : daul@physique.ens.fr},\\
24 rue Lhomond, 75231 Paris CEDEX 05, ~France}.
\end{center}
\vfill
\begin{abstract}
\baselineskip .4 true cm
\noindent
We propose a matrix-model derivation of the scaling exponents of the critical
and tricritical q-states
Potts model coupled to gravity on a sphere. In close analogy with the $O(n)$
model, we reduce the determination
of the one-loop-to-vacuum expectation to the resolution of algebraic equations;
and find the explicit scaling
law for the case q=3.
\end{abstract}

\medskip
LPTENS 94/xx

November 1994
\vfill
\end{titlepage}

%
%   definitions
%

\def \tr { \mbox{Tr}}
\def \trn { {\mbox{Tr} \over N } }
\def \la {\lambda}
\def \wil {W(A_1,A_2)}
\def \pint {\int \!\!\!\!\!\! - \,}
\def \surn {{1 \over N}}
\def \surnn {{1 \over N^2}}
\def \ipi { {1\over 2 i \pi}}
\def \ippi { {1\over {(2 i \pi)^2}}}
\def \dd  {\partial}
\def \limite {\rightarrow}
\def \CO {{\cal O}}
\def \be {\begin{equation}}
\def \ee {\end{equation}}
\def \bea {\begin{eqnarray}}
\def \eea {\end{eqnarray}}
\def \re {\mbox{Re }}
\def \im {\mbox{Im }}
\def \su {\mbox{supp\hskip 1mm }}
\def \gstr {\gamma_{str}}
\def \np{Nucl. Phys. }
\def \pl{Phys. Lett. }
\def \plb{Phys. Lett. {\bf B} }
\def \pr{Phys. Rev. }
\def \prl{Phys. Rev. Lett. }
\def \cmp{Comm. Math. Phys. }

\section{Matrix realization of the Potts model}

 Matrix models have proved to be an efficient means to deal with the
statistical properties of matter coupled
to two-dimensional gravity: instead of performing functional averages on the
metric and embedding fields, we
consider a
matrix-valued field theory whose diagrammatics corresponds to the enumeration
of simplicial complexes
(discretized surfaces) dressed with other (i.e. non-geometrical) degrees of
freedom (e.g. Ising, Potts
spins, a scalar field...).

Here, we shall examine \cite {kazpotts}:
\begin{equation}
Z= \int d\phi \exp \, -N \tr \Big[ \sum_i {m^2 -1 \over 2} \phi_i^2
-\sum_{\langle ij\rangle} \phi_i \phi_j
+\sum_i {\bar{g}\over 3} \phi_i^3 \Big]
\label {eq:partition}
\end{equation} with $\phi_{1 \ldots q} \, \, q \, \, N\times N$ hermitean
matrices, and $\langle i j
\rangle $ denoting any pair of different indices.
Expanding in powers of $g$ and performing the gaussian integral, we obtain
various $\phi^3$ fatgraphs
(we represent any propagator with a double-line and corresponding row/column
indices attached, and any
vertex with a fat Y with the index structure of the trace of a third power
$\phi_{ab}\phi_{bc}\phi_{ca}$)
where each vertex has been attached a number $i=1,\ldots q$ ( which matrix has
been chosen to produce that
vertex?). We can as well consider dual graphs: various gluings of colored
triangles in an orientable
surface without boundary.
As for the statistical weight, we get a contribution from the vertices,
recognized as a cosmological
constant (energy$\sim$area = number of triangles); and from the
propagators,e.g.:
\begin{equation}
{\big\langle \phi_1 \phi_1 \big\rangle \over \big\langle \phi_1 \phi_2
\big\rangle} = m^2 -q+1
\end{equation}
where the energy of a satisfied link (neighbouring triangles with the same
color) re-normalizes the
cosmological constant (there is $1.5$ link per unit area) and the relative
energy of frustrated bonds
produces the required statistical weight of the Potts model; the temperature is
truely positive, provided
that $m^2 > q$. For any $m$, the average size of the surfaces will grow with
$\bar{g}$, and there is a critical value $\bar{g}_c(m)$ near which the
thermodynamic regime
is achieved; if we also tune $m$, the temperature, the spin-correlation
length will also diverge, and we shall obtain the critical regime.

So far, diagrams of all genera appear; but they are weighted by
$N^{\chi = Euler\,  characteristic}$ as is
easily
seen: a factor of N accompanies any vertex; $1/N$ for each propagator; and a
choice of a free index
($a=1,\ldots N$)
for every loop in the $\phi^3$ fatgraph. Taking the logarithm, we obtain
connected diagrams only,
and sphere-like diagrams are extracted in the $N\to \infty$ limit.

In all solved matrix models, a tremendous simplification appears in that
limit: variables are changed
from $\phi$ to $\Omega , \lambda$ according to $\phi = \Omega^+ {\rm
diag}_{1\ldots N} (\lambda) \Omega$
and when the integral over the diagonalizing unitary matrices is preformed, one
is left with an integral
over the $N$ eigenvalues per original matrix with an effective energy $\CO (N
\tr) = \CO (N^2)$,
that is: a semi-classical problem (note that the ground state corresponds
to eigenvalues of order 1).

Indeed, the integral over $\Omega$ is trivial for a one-matrix model where
the energy is invariant under
conjugation. For terms coupling two matrices $$\tr \phi_1 \phi_2 = \tr
\Omega_1^+ {\rm diag}(\lambda_1) \Omega_1
\Omega_2^+ {\rm  diag}(\lambda_2) \Omega_2 $$ the required integral involves
the relative angle $\Omega_1 \Omega_2^+$ and is exactly known \cite {itzub}.
However, when the model couples 3 or more matrices in a cyclic way ($\phi_1
\phi_2 , \phi_2 \phi_3 , \phi_3 \phi_1 $) the relative angles are no more
independent, the semi-classical reduction is no longer possible
and we do not know how to solve such models - that would correspond
to matter with $c>1$ (e.g. a 2-dimensionnal lattice of matrices to produce
surfaces embedded in 2 dimensions).

This problem is only apparent in the case under consideration: the Potts model
on
a flat regular lattice is known to have central charge less than one, and
the cyclic structure of (\ref {eq:partition}) can be disentangled by
introducing
an auxiliary gaussian matrix symmetrically coupled to the Potts matrices
\cite {kazpotts, kostovjaca}
\begin{equation}
Z= \int dX \, d\phi \exp - N\tr \Big[ { \sum_i V(\phi_i)} +X^2 /2 -\sum_i
\phi_i X \Big]
\label{eq:boulatov}
\end{equation}
where $ V(\phi) = {m^2 \over 2} \phi ^2 +{\bar{g}\over 3} \phi^3$.

Now the relative angles $\Omega_{X/\phi_i}$ are independent and can be
integrated out: to solve the problem for planar random graphs, we just have
to determine the distribution of the eigenvalues of the auxiliary and Potts
matrices in the lowest-energy configuration. We shall note $\rho$ the density
of eigenvalues for $X$, that is $\rho(x) \, dx = $ the fraction of eigenvalues
in $[x,x+dx]$; $\sigma$ will correspond to the Potts matrices (we assume no
symmetry breaking, so that all q matrices have the same spectrum at
equilibrium).
The associated resolvents are $f$ and $g$:
\begin{equation}
f(z) = \int {\rho(x) dx \over z-x}
\end{equation}
and correspondingly for $g$; they are holomorphic functions on the complex
plane cut along the supports of their respective densities (finite intervals).
The densities do depend on $m$ and $g$, and scaling laws exist between the
critical exponents attached to them and critical indices of the spin
system: for instance, the string susceptibility $\gstr$ which governs
the area-dependence of the two-point function is known through the exponent
with which the density of eigenvalues vanishes near the edge of its support
{\em exactly} at the critical point $\sigma(y)\sim y^\delta$
\be
\gstr = -(\delta -1)
\ee
 We shall find the densities in the critical regime.

When we change from hermitean to polar variables, a Jacobian
$\Delta(\lambda)^2$
appears: the square of the Vandermonde determinant of the eigenvalues.
This leads to (repulsive) interactions between the eigenvalues, to which we
shall add the effect of the potential dressed by the interactions between
matrices: if we set
\begin{equation}
I(X) = \int d\phi e^{- N \tr \big[ V(\phi) - X.\phi \big]} = I(x_1,\ldots,x_N)
\end{equation}
the dressed potential seen by the $x$'s is ${N\sum x_i^2} -q\log \big(
I(x)\big)$
and the classical equilibrium equation reads
\begin{equation}
{2\over N}{ \sum_{j:j\neq i} {1\over x_i - x_j} }-x_i+q\, w(x_i) =0
\ee
where $w(x_i) ={1\over N} {\dd \over \dd x_i}\log(I) = \langle \phi_{ii}
\rangle $, the
average being taken for a matrix $\phi$ that fluctuates in its own cubic
potential
under the influence of a {\em diagonal} matrix $X$ with spectrum $\rho$.

So, we have
\be
2 \re f(x) -x +q w(x)=0 \, ,\, x\in \su \rho
\label{eq:mvt:f}
\ee
and if we introduce
\be
J(\phi) = \int dX e^{-N \tr \big[ {1\over 2}X^2 - X.\phi\big]} I(X)^{q-1}
\ee
to express the effect on $\phi$ of the fluctuations of the auxiliary matrix
(that feels the influence of the $q-1$ other Potts matrices); writing
\be
\zeta(\lambda_i) ={1\over N}{\dd \over \dd \lambda_i}\log(J) =\langle X_{ii}
\rangle
\ee
we obtain
\be
2 \re g(y) -V^{'}(y)+\zeta(y) =0 \, ,\, y\in \su \sigma
\label{eq:mvt:g}
\ee
 This Hartree-Fock formulation of the large $N$ situation is similar to that
 used in \cite{boulatov:bethe} to find the common distribution of eigenvalues
 of matrices connected along a Bethe tree; here, the tree is finite (with $q+1$
sites), and
 the density is site-dependent. The resolution will proceed along similar
lines,
 using results of \cite{migdal:laformule}.

\section{The critical point}

We can extract useful information without solving explicitly
(\ref{eq:mvt:f},\ref{eq:mvt:g}), just by using analytic properties
of generalized resolvents. Let us introduce
\be
W_k(z)={1\over I} \Big({1\over z-{1\over N}\dd_X}\Big)_{kk} . I = \Big\langle
\Big( {1\over z - \phi} \Big)_{kk}\Big\rangle
\ee
which we shall also write $W_x(z)$, $x$ being the $k$-th eigenvalue of $X$, to
which $\phi_{kk}$ is coupled.
We have \cite{migdal:laformule}
\be
{1\over N} \sum_j {W_k(z)-W_j(z) \over x_k - x_j} = - \surn \Big({d_XI\over
I}.\langle{1\over z-\phi}\rangle
\Big)_{kk}+\langle{\phi
\over z-\phi}\rangle_{kk}
\label{eq:migdal:W}
\ee
This formula can be obtained by considering the contracted differential
$d_{ab}f_{bc}$ of the matrix-valued function:
$$f:\, X\mapsto \langle{1\over z-\phi}\rangle $$
(\ref{eq:migdal:W}) only expresses the equivariance of $f$
$$f(\Omega X\Omega^+)=\Omega f(X) \Omega^+ $$
In the large $N$ limit, (\ref{eq:migdal:W}) reduces to
\be
\int dx \, \rho(x) {W_a(z) -W_x(z) \over a-x} = -w(a)W_a(z) +z W_a(z) -1
\label{eq:w}
\ee
Similarly, we define
$$Z_k(z) ={1\over J} \Big({1\over z-{1\over N}\dd_\phi}\Big)_{kk} . J =
\Big\langle \Big( {1\over z - X} \Big)_{kk}\Big\rangle $$
and we introduce the generalized resolvents
\begin{eqnarray}
F(z,s)&=&1-\int dx{\rho(x)W_x(z)\over s-x} \nonumber \\
G(z,s)&=&1-\int dy {\sigma(y) Z_y(z) \over s-y}
\label{eq:resol}
\end{eqnarray}
Expliciting $F$ we have
\begin{eqnarray}
F(z,s)&=& 1-\surn \sum_k {1\over s-x_k}\Big\langle \Big({1\over
z-\phi}\Big)_{kk}\Big\rangle \nonumber \\
      &=& 1-\surn \Big\langle \tr \, {1\over s-X}\, {1\over z-\phi} \Big\rangle
\label{eq:W:simple}
\end{eqnarray}
and, as $X$ and $\phi$ change roles from $F$ to $G$, $G(s,z)=F(z,s)$.

For large $z$, $W_{\cdot}(z)$ is well defined and $F(z,\cdot)$ has a cut along
$\su \rho$:
\begin{eqnarray}
F(z,a\pm i \epsilon)&=& 1-\pint \, dx {\rho(x) W_x(z) \over a-x} \pm i \pi
\rho(a) W_a(z) \nonumber \\
                    &=& W_a(z) \big[ z-w(a) -\re f(a) \pm i \pi \rho(a)\big]
\end{eqnarray}
in virtue of (\ref{eq:w}). We assume that $w$ and $\zeta$ can be analytically
continued near the supports of $\rho,\sigma$
(independently below and above the cuts: we shall see that the edges are
branching points)
and we set
\be
u=w+f \, , \, v=\zeta +g
\ee
so that
\begin{eqnarray}
F(z,a) &=& W_a(z) (z-u(a)) \nonumber \\
G(z,b) &=& Z_b(z) (z-v(b))
\end{eqnarray}
and, as these functions go to one at infinity (\ref{eq:W:simple}), we can write
a dispersion integral for their phases
\begin{eqnarray}
F(z,s)&=& \exp \, -\oint_{\su\rho} {d\tau \over 2 i \pi}{1\over \tau -s}\log
\big[ z-u(\tau)\big] \nonumber \\
G(z,s)&=& \exp \, -\oint_{\su\sigma} {d\lambda \over 2 i \pi}{1\over \lambda
-s}\log \big[ z-v(\lambda)\big]
\end{eqnarray}
with contour integrals going counterclockwise along $\su \pm i \epsilon$, this
being a valid representation
for large $s,z$. (Proof: for fixed large $z$, the ratio of both sides is a
holomorphic function of $s$, with no cut,
going to 1 as $s\rightarrow \infty$, thus equal to 1.) We note $[\alpha,\beta]
= \su \rho$.

Let us express the symmetry property: $G(s,z)=F(z,s)$: we change the variable
in the first contour integral from $\tau$
to $u(\tau)$ and then integrate by parts to obtain
\be
F(z,s) = \exp \, \oint_{u({\cal C})} {du \over 2 i \pi} {\log \big[\tau(u)
-s\big] \over u-z}
\ee
To be precise, the former integral involved $u_\pm$ along $[\alpha,\beta]\pm i
\epsilon$; the new expression involves
the integrals along two arcs, joining $u_{+}(\alpha)=u_{-}(\alpha)$ (for $\rho
(\alpha)=0$) to $u_{+}(\beta)= u_{-}(\beta)$, of
the corresponding inverse function $\tau$: see Figure~1.
\begin{figure}
 \epsfysize= 5cm \epsfbox{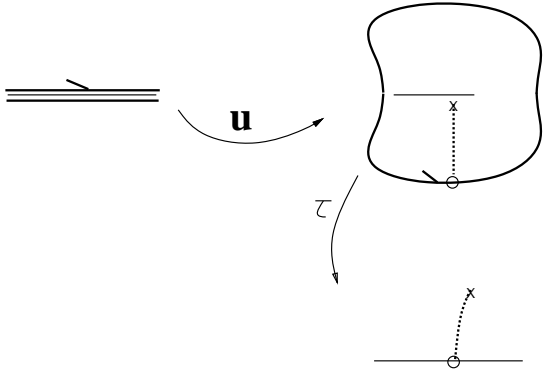}
 \caption{In $u\circ \tau = id$, we evaluate $u$ on the sheet on which lies the
value taken by $\tau$ after analytic
continuation (dashed line).}
\end{figure}

We note that the upper segment is mapped onto the lower arc because $\im u_{+}
<0$: so, the orientation of the image contour
depends on the order of $u(\alpha),u(\beta)$.

We expect as few singularities as possible for $u$ and $\tau$,
because we introduced the smallest possible number of coupling constants
(should we have taken more, we could have produced
higher order criticality through a richer singularity structure): we assume we
can flatten the integration contour to
$\big(u(\alpha),u(\beta) \big)$ and analytically continue $\tau$ without
encountering any singularity, so that
\begin{eqnarray}
 F(z,s) =&  \exp \, \oint_{(u(\alpha),u(\beta))} {{du \over 2 i \pi}{1\over
u-z} \log \big[\tau(u) -s\big] }
                                   \nonumber \\
          = & G(s,z)= \exp \, -\oint_{\su\sigma} {{d\lambda \over 2 i
\pi}{1\over \lambda -z}\log \big[ v(\lambda)
-s\big]}
\end{eqnarray}
We knew that equality for large $z,s$: under that form, contour integrals are
seen to be equal (up to $2 i \pi$, but they
vanish at infinity!) for large $s$ and any non-real $z$. Considering the
discontinuity on the real $z$-axis, we obtain for
large $s$:
\be
\chi_{\scriptstyle \lambda \in \big(u(\alpha),u(\beta)\big)}\log
\Big[{\tau_{+}(\lambda)-s \over \tau_{-}(\lambda)-s }\Big] =
\pm \,  \chi_{{\scriptstyle \lambda \in \su \sigma}}  \log
\Big[{v_{+}(\lambda)-s \over v_{-}(\lambda)-s }\Big]
\ee
where the sign depends on the relative order of $u(\alpha),u(\beta)$. We
conclude that $(u(\alpha),u(\beta))$ covers $\su
\sigma$ and that
\be
\tau_{\epsilon} (\lambda) = v_{\pm \epsilon} (\lambda) \, , \, \lambda \in \su
\sigma
\ee
For instance, if $u$ preserves the order of $\alpha,\beta$: $\tau_\pm = v_\pm$
on $\su \sigma$ and
\be
\lambda = u_{+}\big[v_{-}(\lambda)\big] \ee
for $\lambda \in \su \sigma$ and so, for any $\lambda$ by analytic
continuation. In any case, attention shall be paid to
determine the u-sheet on which $v(\lambda)$ has to be evaluated: one has to
follow the value of $\tau$ as its argument moves
from $u({\cal C})$ towards the real axis.

This inversion relation $u\circ v = id$ always holds, in the planar limit,
whether we are at the critical point or not.
However, if $v$ is singular at $\lambda$ then $u$ is at $v(\lambda)$: but at
the critical point, any function has
all its singularities lying at the same place (see section
\ref{section:scaling}), and we expect each density to develop
a singularity at one edge of its support; so, these singular edges are
exchanged under $u,v$.

 We call $\gamma,\delta$ the exponents with which $\rho,\sigma$ vanish at the
singular edges: they govern the corresponding
leading singularities of $f,g$. For example, if $\rho(\alpha +\epsilon) = cst
\, \epsilon^\gamma + \ldots$ we have
$$ f(\alpha+\zeta) = regular \, \, + {(cst >0)\over \sin \pi\gamma }
(-\zeta)^\gamma +\ldots $$
where the $\gamma$-th power of a positive number is taken positive. Using
(\ref{eq:mvt:f},\ref{eq:mvt:g}) we obtain similar
expansions for $u,v$, e.g. (singularities at the left edge)
\bea
u_\pm (\alpha +z)&=& reg \, + C \sin \pi \gamma \, \big[{q-2\over q} \cos \pi
\gamma \mp i \sin \pi \gamma \big] z^\gamma +\ldots \nonumber \\
v_\pm (u(\alpha)+z)&=&reg \, +C' \sin \pi \delta \, \big[ - \cos \pi \delta \mp
i \sin \pi \delta \big] z^\delta +\ldots
\label{eqa:dvlpt}
\eea
with positive constants $C^{(')}$. Moreover, we expect $1<\delta<2$ (see
\cite{plan:diag}, \cite{dkk}, and section
\ref{section:scaling}).

For that singularity to disappear in $u\circ v = id$, $\gamma$ has to be equal
to $\delta$ or to its inverse; in the latter
case, the linear term in the regular part of $v$ necessarily vanishes. In the
first case, we write that some linear
combination of $C,C'$ vanishes and obtain a relation between the phases of the
bracketed terms of (\ref{eqa:dvlpt}).
In the other case ($\gamma \delta = 1$), the requirement that $z=
(\cdots)(z^\delta)^\gamma$ also gives a relation
between these phases.

Investigating all the possible situations (respective positions of the singular
edges, relation between $\gamma$ and $\delta$)
we discover that, for singularities occuring both at the same edge:
\bea
{\mbox{either}}&\gamma = \delta^{-1}&\mbox{and } \cos \pi (1- 1/\delta)
=\sqrt{q}/2 \nonumber \\
 \mbox{or}     &\gamma = \delta     &\mbox{and } \cos \pi (\delta -1)
=\sqrt{q}/2 \nonumber
\eea

The case when one singularity occurs at the left edge and the other on the
right edge, appears to be inconsistent: this is
related to our choice of sign of the $X.\phi$ term in (\ref{eq:boulatov}); with
that choice $X$ is localized near
$\sum_i \phi_i$,
and we understand that the spectra become simultaneously singular at
corresponding edges.

Expressing our results in terms of $m =-1/\gstr$ (parametrization of $m,m+1$
unitary models) we find
\bea
\gamma =\delta^{-1}&\mbox{and }\sqrt{q}/2 = \cos \pi/(m+1) \nonumber \\
\gamma =\delta &\mbox{and } \sqrt{q}/2 = \cos \pi/m
\eea
and recognize the (flat space - regular lattice) critical and tricritical Potts
model \cite{cardy}.
In particular, we recover the unitary (5,6) model as the critical 3-states
Potts model. This result shows that
it is important not to confuse between the exponents of $\rho$ and $sigma$,
otherwise one fails to identify
the critical Potts model.

Before turning to the study of the vicinity of the critical point, a last
remark is in order: we said that the existence
of a singularity for $u$ at $\la$ implied a singularity for $v$ at $u(\la)$.
This is true exactly at the critical point,
but not in its neighbourhood: out of the critical point, densities of
eigenvalues vanish as square-roots at the ends of their
supports (as in the simple one-matrix case \cite{plan:diag}; see also the
scaling functions in the next section). Then,
$u(\alpha +z)$ is an invertible power series in $\sqrt{z}$, and the condition
on the inverse function is a zero of its
derivative $v'(u(\alpha))=0$. When we approach the critical point, that zero
disappears at the place where two cuts merge
(consider the example of $\sqrt{\epsilon^2 -x^2}$ as $\epsilon \limite 0$) and
the fractional order of the zero of
the density at the edge
of the support is increased: this will be transparent with the more explicit
results of the following section.

\section{The scaling regime}
\label{section:scaling}

For any values of the temperature and the cosmological constant, we can use
known results about the external
field problem (Kazakov and Kostov, in \cite{kostovjaca}; see also
\cite{gross:newman}) to compute exactly
$w$ in the large $N$ limit: setting $\alpha = 1/m$, $g=\bar{g}/m^3$,
\be w(x) /\alpha = {\sqrt{\alpha x+c}\over \sqrt{g}} -{1\over 2g} +{1/2}\int
{\rho(t) \, dt \over \sqrt{
\alpha t+c}} {1\over \sqrt{\alpha x+c} +\sqrt{\alpha t+c}} \ee
$w(x)$ being the restriction of a holomorphic function with a semi-infinite cut
$(-\infty$,$-c]$ whose
location, at the left of the support of $\rho$, is given by
\be c+\sqrt{g} \int {\rho(x) \, dx \over \sqrt{\alpha x+c}} = {1\over 4g} \ee
(for positive $g$, which we shall assume is the case; for negative $g$, the cut
is located to the right
of the support of $\rho$)

The eigenvalues of $X$ are in the lowest-energy configuration when
\be x = 2 \pint {dy \, \rho(y)\over x-y} + q \alpha \Big({\sqrt{\alpha
x+c}\over \sqrt{g}} -{1\over 2g}
+{1/2}\int {\rho(t) \, dt \over \sqrt{\alpha t+c}} {1\over \sqrt{\alpha x+c}
+\sqrt{\alpha t+c}}\Big) \ee
on the support of $\rho$. Changing notations to
\be u = \sqrt{\alpha t+c}, \pi(u) =\rho(t) \ee and introducing
\be G(z) = \int {du \, \pi(u) \over z-u} \ee
the equilibrium equation reads
\be \label{eqdebase} 2 \re G(z) + (2-q) G(-z) = P(z) \ee
with
\be P(z) = {z^2 -c \over \alpha} -{q \alpha \over \sqrt{g}} z +{q \alpha \over
2g} \ee

Note that $\pi$ is not normalized, but
\be {1\over 4g} = c+ {\alpha \sqrt{g} \over 2} \int \pi \ee
while the normalization of $\rho$ now reads
\be \int u \pi(u) \, du = \alpha /2 \ee

When $q=2$ this equation is easily solved by a dispersion integral and we
recover the known results about
the Ising model on a dynamical lattice, with
\be \alpha_c^2 = {\sqrt{7}-1\over 12}, g/\alpha^3 = \sqrt{10} \ee
($\alpha_c < .5$ effectively corresponds to some positive temperature), and the
usual scaling law for
the resolvent (one-loop function).

For $q=0$ or 4, (\ref{eqdebase}) has a $Z_2$ symmetry ($z\rightleftharpoons
-z$) and its solution
is expressed through elliptic integrals; we recall that $q \to 0$ corresponds
to the statistics of
tree-like polymers on a random lattice and has been solved in \cite{kazpotts}.

In the general case, the critical point is reached when the support of $\pi$,
$[a,b]$, fuses with
its mirror image ($a \to 0$), that is, when the singularity of $w$ reaches the
support of $\rho$. The
scaling behaviour of $G$
\be G(a \cosh \theta) \sim a^{(\ldots)} g(\theta) \ee
is imposed by (\ref{eqdebase})
\be g(\theta) + g(\theta+2 i \pi) +(2-q) g(\theta+i \pi) =0 \ee
that is
\be g(\theta) = cst \, \cosh \kappa \theta \ee
with $\cos \kappa \pi = q/2 -1$ which corresponds to the exponent found in the
first section for the
two-critical Potts model.

So, we found the scaling behaviour of the resolvent of $\pi$; to obtain the
one-loop function for a Potts
matrix, we shall compute the resolvent of $\rho$ and use the inversion formula.
To do so, however, we need
to know the position of the support
of $\rho$ which involves $c$: and this parameter is left undetermined when we
extract the singular
part of $G$ in the scaling limit. To assert a definite conclusion about $\rho$
(and not $\pi$) we need
compute $c$: that we did for $q=3$, an unsolved model up to now.

A general way to proceed with (\ref{eqdebase}) \footnote{We thank B.Eynard for
his advice} is to shift $G$
by the polynomial $Q$ that satisfies $2 Q(z) +(2-q) Q(-z) = P(z)$, to obtain a
homogeneous equation for
$f= G-Q$. Introducing $\theta$ such that $\cos \theta = 1-q/2$ ($\theta = 2\pi
/3$ for $q=3$), we define
\be \phi_\pm (z) = f(z) + e^{\pm i \theta } f(-z) \ee

These holomorphic functions have a double cut $[-b,-a]$, $[a,b]$, are related
by $\phi_-(z) = e^{-i\theta}
\phi_+(-z)$. Through the right cut, $\phi_+$ is continued into $-\phi_-$; and
through the
left cut, into $-e^{2i\theta}\phi_-$ according to (\ref{eqdebase}).

When $\theta$ is commensurable to $\pi$, we can raise $\phi_\pm$ to some
integer power so as to obtain
holomorphic functions which realize a double-cover of the sphere (with two
cuts). Namely, when $q=3$, we set $\psi_\pm =
\phi_\pm^3$, so that $\psi_-(z) = \psi_+(-z)$ and the analytic continuation of
$\psi_+$ through any cut
leads to $-\psi_-$. At infinity, $\psi$ behaves as $z^6/\alpha^3$. We obtain a
meromorphic function
on the torus (double cover of the two-slit plane) which can be written
\be \psi_\pm = \alpha^{-3} \big( \sqrt{(z^2-a^2)(z^2-b^2)}S(z) \pm i R(z) \big)
\ee
with some monic even polynomial $S$ of degree 4; and an odd polynomial $R$ with
degree $\leq 5$.

The vanishing of the sum of the residues of $\psi'/\psi$ on the torus shows
that $\psi$ has twelve zeroes.
These have to be triple zeroes, since $\phi = \psi^{1/3}$ has no branching
point and we conclude that
\be \label{eqcarre}
S^2(z) (z^2-a^2)(z^2-b^2) + R^2 (z) = (z^2 -\lambda^2)^3 (z^2 -\mu^2)^3 \ee

This requirement allows us to solve the Potts model: for given $\alpha$ and
$g$, we look for $a,b$ and the
coefficients of $R,S$ (3 and 2 parameters); (\ref{eqcarre}) gives 4 constraints
(for a sixth-order polynomial
(in $z^2$) to have two triple roots), and 3 other equations are given by the
asymptotic expansion of $f$
(coefficient of $z$; eliminate $c$ between the coefficients of $z^0$ and
$z^{-1}$; coefficient of $z^{-2}$).

Precisely at the critical point, we know that $a=0$ and $\rho(t_c +dt) \sim
dt^{5/6}$ so that $\pi(u)
\sim u^{5/3}$:
inspection of \ref{eqcarre} then shows that 0 is a zero of $S$ with
multiplicity 4, and with multiplicity 5
for $R$. We thus have identified all coefficients in \ref{eqcarre}
\be  (z^4)^2 z^2 (z^2-b^2) + (b z^5)^2 = (z^2)^3 (z^2)^3 \ee
and we find that
\be \alpha_c^2 = {\sqrt{47}-3 \over 38} , g_c = 27 \alpha_c^4 \sqrt{{665 \over
486 (\sqrt{47}-3)}} \ee

We then solve (\ref{eqcarre}) in the vicinity of that point to investigate the
scaling limit; to obtain
a generic perturbation, we can shift $g$ while keeping $\alpha = \alpha_c$. We
find the following behaviours
\be db \sim a^2 , \la \sim \mu \sim a^{5/6} \mbox{ with } \la^2 +\mu^2 \sim a^2
\ee
the roots of $R,S$ scale as $a$, while the leading coefficient of $R$ varies
with $a^{10/3}$, just like $g$.
And $dc = a^2/2 + \ldots $

For the scaling behaviour of $f$ we get
\bea
 f(z=a\cosh y + i0) = -{i\over \alpha \sqrt{3}}& \Big[ e^{-2i \pi /3}\big( i
a^5 b & [ \sinh y (\cosh^4 y
-{3\over 4} \cosh^2 y +1/16) \nonumber \\
 & & - \cosh y (\cosh^4 y -{5/4}\cosh^2 y +5/16) ]\big)^{1/3}  \nonumber \\
  &- e^{2i \pi /3} \big( i a^5 b & [ \sinh y (\ldots) + \cosh y (\ldots)]
\big)^{1/3} \Big] \eea
where the branch of the third root is fixed by continuation from the behaviour
of $f$ at infinity, and where
the bracketed combinations of hyperbolic lines reduce remarkably to $-\exp
(-5y/16)$ and $\exp (5y/16)$
so that
\be f(a \cosh 3y +i0) \sim a^{5/3} \cosh 5 (y -{2i\pi /15}) \ee

The explicit limit we found for $c$ allows us to see that $\rho$ also has a
scaling limit expressed
with hyperbolic lines because $u\sim a\cosh 3y$ corresponds effectively to $dt
\sim a^2 \cosh 6y$
(the coefficient $1/2$ in $dc = a^2/2 +\ldots$ is really important here)

Finally, using $v_-\circ u_+ = id$ we find $v_-$, then $g$, the one-loop
function for a Potts matrix (partition
function of a disk with a uniform color along the boundary)
\be g(\epsilon^5 \cosh 5t) = \mbox{ regular }+ cst \, \epsilon^6 \cosh 6(t-i\pi
/5) + \ldots\ee

Such scaling laws are universal among unitary models: in \cite{dkk} however, a
two-matrix formulation was used
to produce any $(m,m+1)$ model, with multicritical points analogous to
Kazakov's one matrix-model (the
geometry of the different types of polygons used to build the surface -not only
triangles- conspires
with the Ising spins to produce different conformal models) and the
identification with the explicit
model here discussed is rather unclear. We can just argue that in both cases,
we compute averages
of macroscopic insertions of the identity operator.

To conclude, we would like to consider again the 3-critical point: how could it
appear in the first section,
while we missed it in the above exact computation? To reach tricriticality, we
need two coupling constants
for the statistical model, and not only the temperature: it seems that we would
obtain a model in the
right universality class if we added an $\tr X^3$ term to the action, to
produce a kind of dilute Potts gas;
this modification would not change the lines of the first section, where the
simplest type of criticality
was assumed with given number of parameters, while it requires a new
investigation of the derivation given
in the last section. That seems interesting to clarify, in order to gain a
better understanding of the
$q \to 4$ limit, where these two RG fixed points merge into a first-order fixed
point, this limit
being the $c\to 1$.

\end{document}